\documentclass[prl,amsmath,amssymb,floatfix,superscriptaddress,notitlepage]{revtex4}
\usepackage{graphicx,enumitem}
\usepackage{hyperref}
\usepackage{bm}
\usepackage{amssymb}
\usepackage{xifthen}
\usepackage{color}
\usepackage{datetime}
\usepackage{float}
\usepackage{bm}
\usepackage{changepage}
\usepackage{bbold}
\usepackage{multirow}
\usepackage{tikz}
\usetikzlibrary{positioning}
\usepackage{adjustbox}

\newcommand{\ee}[1]{\begin{align} #1 \end{align}} 								
\newcommand{\vc}[1]{\vec{\mathbf{#1}}} 										
\newcommand{\intg}[2]{\int \! \mathrm{d}^{#1}\vc{#2} \ }							
\newcommand{\nn}[1][]{\ifthenelse{\isempty{#1}}{\nonumber \\}{\nonumber}}			
\newcommand{\dv}{\partial }												
\newcount\colveccount													
\newcommand*\colvec[1]{
        \global\colveccount#1
        \begin{bmatrix}
        \colvecnext
}
\def\colvecnext#1{
        #1
        \global\advance\colveccount-1
        \ifnum\colveccount>0
                \\
                \expandafter\colvecnext
        \else
                \end{bmatrix}
        \fi
}
\newcommand{\image}[3]{
	\begin{figure}[htbp]
		\includegraphics[width=\linewidth]{#1}
		\caption{#2}
		\label{#3}
	\end{figure}
} 															

\tikzstyle{mybox} = [ fill=blue!5, inner sep=10pt, inner ysep=20pt]
\tikzstyle{boxtitle} =[fill=blue!20, text=black]

\begin{document}

\title{Information Dynamics at a Phase Transition }

\author{Damian Sowinski}
\email{Damian.Sowinski.GR@dartmouth.edu}
\affiliation{Department of Physics and Astronomy\\ Dartmouth College,Hanover, NH 03755, USA}

\author{Marcelo Gleiser}
\email{Marcelo.Gleiser@dartmouth.edu}
\affiliation{The Department of Physics and Astronomy\\ Dartmouth College,Hanover, NH 03755, USA}

\date{Last Updated: \today, \currenttime}

\begin{abstract}
We propose a new way of investigating phase transitions in the context of information theory. We use an information-entropic measure of spatial complexity known as {\it configurational entropy} (CE) to quantify both the storage and exchange of information in a lattice simulation of a Ginzburg-Landau model with a scalar order parameter coupled to a heat bath. The CE is built from the Fourier spectrum of fluctuations around the mean-field and reaches a minimum at criticality. In particular, we investigate the behavior of CE near and at criticality, exploring the relation between information and the emergence of ordered domains. We show that as the temperature is increased from below, the CE displays three essential scaling regimes at different spatial scales: scale free, turbulent, and critical. Together, they offer an information-entropic characterization of critical behavior where the storage and processing of information is maximized at criticality.
\end{abstract}

\maketitle


\section{Introduction}
\noindent
As a mathematical discipline, information theory had its origins in the fairly recent past, under Claude Shannon's seminal work \cite{Shannon}. Shannon proposed an entropic measure of information for message transmission and decoding, designed originally for sets of $A$ symbols (e.g. an alphabet) $\mathcal A = \{a_1, a_2,...,a_A\}$ each appearing with probability $p_i=p(a_i)$,  in messages of finite length $M$. Since we must ask $A$ yes or no questions to know which symbol $a_i$ is being stored, the number of bits to store a symbol is $\log_2 A$. 
It follows that to store a message with $M$ symbols, the number of bits is simply $M\log_2 A$. However, alphabets are often compressible, in the sense that certain letters are more probable than others in a given message. This can be quantified using Huffman encoding \cite{Huffman}, where the amount of information used to encode a symbol $a_i$ in an alphabet $\mathcal A$ is $I(a_i) = -\log_2 p_i$: very improbable symbols carry a lot of information. The average amount of information in an alphabet $A$ is then given by Shannon's information entropy
\ee{
\langle I \rangle \equiv S_S[\mathcal A] = \sum_{a_i\in \mathcal A} p_i I(a_i).
}

\noindent
This expression is maximized at $S_S= \log_2 A$, when all symbols have the same probability of occurrence, $p_i=1/A$. In Shannon's interpretation, an alphabet that maximizes entropy is incompressible: the set of symbols cannot be encoded with less information. When entropy is maximal, ignorance is also maximal in the sense that all symbols are equiprobable. Compressibility, then, is a measure of the departure from this maximum entropic value. The more compressible an alphabet, the lower its information entropy. Below, we will establish an analogy between compressibility in an alphabet and the relative probability of different momentum-modes in a field configuration.

It didn't escape physicists that Shannon's formula is very close to the Gibbs thermodynamic entropy for a classical system with a discrete set of microstates $\phi_i$ with energy $E_i$ and probability distribution $p(\phi_i)$ \cite{Sethna},
\ee{
S_G[\Phi]=-k_B\sum_{\phi_i\in \Phi} p(\phi_i)\ln [p(\phi_i)],
}
\noindent
where $k_B$ is Boltzmann's constant. In the 1950s, Seeger \cite{Seeger},  Brillouin \cite{Brillouin}, Jaynes \cite{Jaynes}, and others \cite{Leff}, attempted to bridge information theory and statistical physics. Brillouin coined the term {\it negentropy} as the information gained by measuring physical properties of a system: the more information, the less uncertainty. Using maximal-entropy inference, Jaynes showed that thermal equilibrium is the state of maximum ignorance, with all energy-equivalent microstates being equiprobable. Since then, much progress has been made toward a connection between information theory and non-equilibrium thermodynamics, including some experimental validation, although a definition of a physically meaningful entropy for generic non-equilibrium states remains elusive \cite{Parrondo}. 

In this work, we explore a novel approach to incorporating information theory to physical systems. The essential idea is to take advantage of the spectral information encoding the {\it shape} of a physical system, where by shape we mean the square-integrable or periodic mathematical function of spatial coordinates describing the system and its dynamical evolution, usually solutions of a $d$-dimensional partial differential equation. 
This shape information is described by a quantity known as {\it configurational entropy} (CE), $S_C[\Phi]$, introduced first in Ref. \cite{GS1} when it was applied to solitons in field theories. Since then, CE has been used to describe a variety of physical systems, from stellar stability \cite{GD1,GN} to the dynamics of symmetry breaking \cite{GS2}, inflationary cosmology \cite{GGS,GG} and, more recently, to studies of critical phenomena \cite{GD2}. 

These previous works were direct applications of the CE formalism, showing phenomenologically  that: i. in all cases considered, the CE was minimized for spatially-localized solutions of the equations of motion \cite{GS1}; ii. CE can detect instabilities in spatially-bound systems \cite{GD1,GN}; iii. CE can be used as a measure of spatial complexity, picking up structure, in particular, coherent structures emerging against noisy backgrounds \cite{GS2} including cosmological contexts \cite{GGS,GG}. (For this one uses the Kullback-Leibler divergence, comparing two distribution functions.) However, these works didn't attempt to develop a more in-depth connection between information theory and spatial complexity, something we hope to remedy here.

Our starting point is the work of Ref. \cite{GD2}, where we presented a preliminary analysis of the behavior of CE during a second-order phase transition \cite{Goldenfeld,Landau}. There, it was noted that the CE appeared to go through a minimum at the critical point, and that the approach to criticality seemed to be characterized by a scaling behavior with the same signature of Kolmogorov turbulence in fluids \cite{Kolmogorov}. However, we fell short of obtaining the proper scaling behavior near and at criticality, and, for this reason, were unable to offer an interpretation of the results based on information theory. We will show that a detailed analysis of the behavior of the CE near and at criticality offers a new way of interpreting critical phenomena in light of information theory. In particular, we will argue that the entropic minimum at criticality corresponds to a state of {\it maximal information storage}, which we will interpret as an increase in the order parameter's compressibility in information space, while an analysis of the various scaling relations we found will allow us to quantify how information flows between different momentum modes as the system passes through its critical point. Taken together with previous results, our analysis clarifies why CE is a general measure of information storage, while the CE density is a measure of information transfer.

\section{The Model and its Numerical Implementation}

Our starting point is the well-known coarse-grained Landau free energy \cite{Landau,Goldenfeld},

\ee{
\label{FreeEnergy}
F[\phi]=\int{d}{\bf x}\left[\frac{1}{2}(\nabla\phi)^2+\sum_{n=1}^\infty a_{n}(T)\phi^{n}\right],
}
\noindent
where the first term represents the energy stored in the field gradients of the order parameter, here taken to be a simple scalar field, whereas the latter is a temperature-dependent effective potential. This functional form describes first and second (continuous) order phase transitions, having been used, with modifications, in a wide variety of systems in the Ising universality class, from binary fluid mixtures \cite{Langer} to symmetry breaking in high-energy physics and cosmology \cite{cosmo}.

Given that our goal is to explore the connection between critical phenomena and information theory, it is enough to work in two spatial dimensions. We will further simplify the Landau model and consider only quadratic and quartic terms, thus limiting our analysis to second order transitions. Our strategy will be to start at a low-temperature ordered state and then drive the system to a different equilibrium state by increasing the temperature. This can be accomplished by using a Langevin equation that models the zero-temperature system in contact with a heat bath at temperature $T$, 
\ee{
\ddot\phi =& -\gamma \dot \phi + \nabla^2\phi + \phi - \phi^3 + \sqrt{2\gamma T} \xi,
}
where we have scaled the dimensionful fields and coordinates $\phi\rightarrow \frac{a_2(0)}{a_4(0)}\phi$ and  $x^\mu \rightarrow \frac{1}{a_4(0)} x^\mu$. The viscosity, $\gamma$, and the random variable, $\xi$, describe the coupling to the heat bath in accord with the fluctuation-dissipation theorem \cite{Sethna}.  $\xi$ is drawn from a normal distribution with zero mean and unit variance, $\langle\xi\rangle = 0$ and $\langle\xi\xi'\rangle=\delta(x^\mu)$.

A phase transition occurs at the critical temperature $T_C$, above which the $\mathrm Z_2$ symmetry holds. To implement the model numerically we use a staggered leapfrog approach with periodic boundary conditions on a square lattice of size $L=512$ and lattice spacing $\Delta x=0.25$. This large and fine lattice allows for a detailed description of the equilibrium states with small error. The lattice is evolved in time steps small enough to satisfy the Courant condition, $\Delta t=0.10$. The spatial and temporal discretization rescale the stochastic term so that the coefficient is now $\sqrt{2\gamma T/\Delta t\Delta x^2}$, as described  in \cite{GB}. The Laplacian is implemented using the maximally rotationally invariant convolution kernel \cite{Lindeberg} with error $\mathcal{O}(\Delta x^2)$. We probe $70$ temperatures in the range $[.05,.95]$ with equal spacing except in the range of the critical temperature. For robust statistics we use an ensemble of 128 lattice simulations at each temperature.

For a given temperature $T$, the field is initialized at $\phi = 1$ and $T_0= 0$. $T_0$ is raised to $T$ in 1500 time steps along a logistic temperature trajectory centered at the 375$^{th}$ time step. This trajectory was chosen so that approximately half the time steps are used to heat the system adiabatically, while the other half insure it reaches equilibrium. Equilibrium is verified by checking equipartition, $\langle\dot\phi^2\rangle=T$.  This method differs from \cite{GD2} in that close to criticality the slow raising of the temperature keeps the ratio of gradient to kinetic energy high enough to prevent any spilling into the opposite vacuum state, minimizing the formation of topological structures and providing a more accurate estimate of the critical temperature. We have checked our results using standard lattice-scaling techniques \cite{Goldenfeld}.

The first and second moments, $\bar\phi$ and $\bar{\phi^2}$, are calculated for each simulation, and statistics are run on the ensemble. The first moment is the mean field, while the second is used to find the susceptibility,
\ee{
\chi(T) = \frac{1}{T}(\bar{\phi^2}-\bar\phi^2).
}
Ensemble statistics for both are plotted in Fig.\boxed{\ref{MeanSusc}}. The inflection point in the mean field and the maximum in the susceptibility give us confidence that the phase transition occurs in the range of temperatures $T_C\in [.45,.483]$, so that we estimate $T_C=.467\pm.033$. These results differ slightly from those of Ref. \cite{GD2}, due to their higher accuracy. This difference will be key to the configurational-entropic analysis to follow. First, we briefly review the formalism of CE, carefully defining how it is to be applied to critical phenomena.

\image{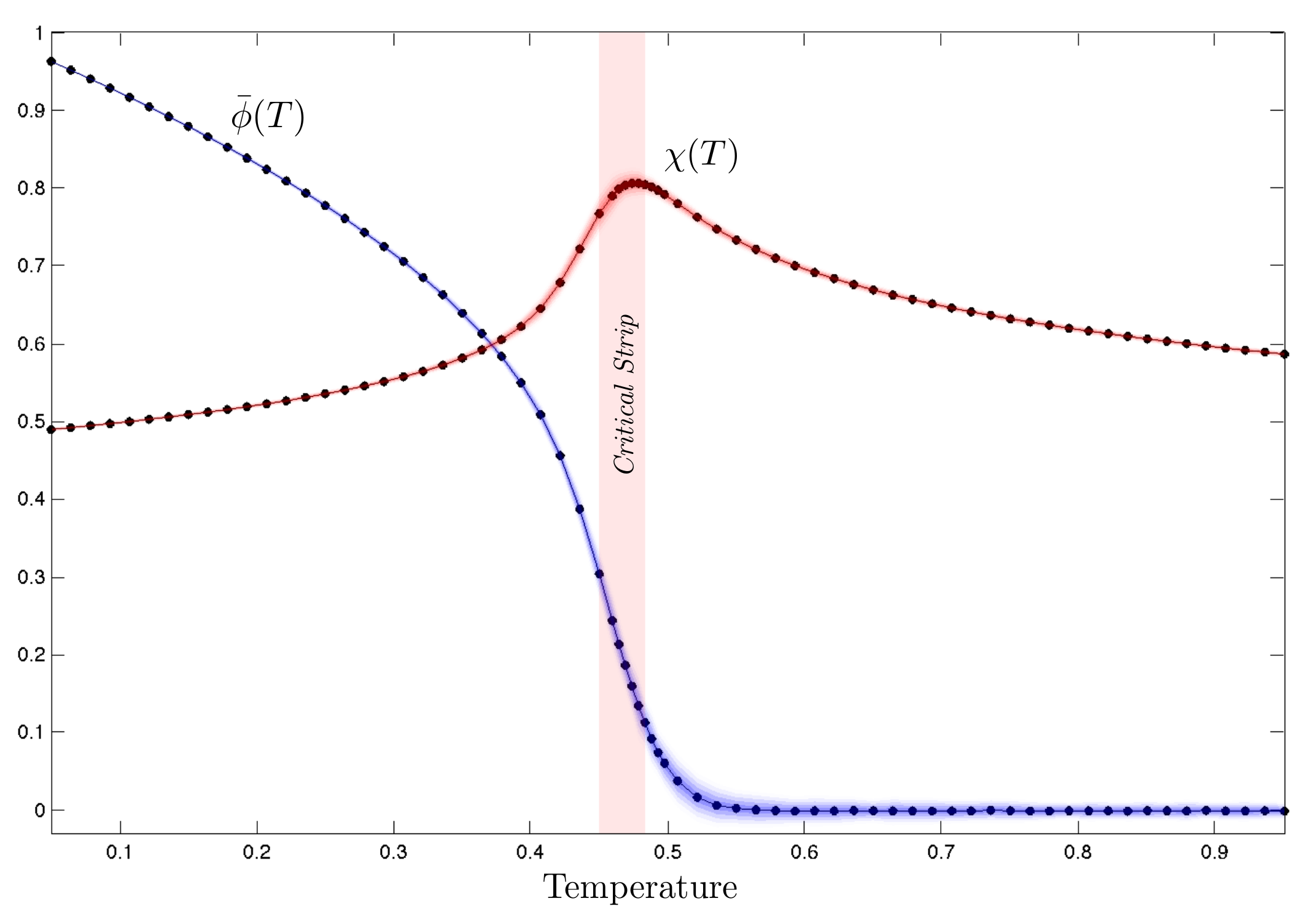}{The ensemble-averaged mean field and susceptibility plotted versus temperature. Color bars are 5$\sigma$ deviations from the ensemble average. The critical strip contains the critical temperature within the accuracy of the simulation.}{MeanSusc}

\section{Configurational Entropy}

To optimize the connection between epistemic states (information about the system from measurements) and ontic states (what the system is actually doing) we must relate the information content of the system at different equilibrium states. A proper account would identify the mechanisms of {\it information storage} and {\it information transfer}. To do so in the context of phase transitions, one must use a probability distribution that reflects the system's self-organization at different spatial scales. The probability distribution of fluctuations about the spatially-averaged mean (cf. Fig. \ref{MeanSusc}) is a potential candidate, since it changes appreciably over a wide array of temperatures. Unfortunately, this distribution leads to a monotonically increasing information measure, without a particularly clear signature at criticality: fluctuation amplitudes do not carry information about the spatial organization of the system, and criticality is characterized by long-distance spatial correlations.   In order to capture the informational properties of spatially-extended domains, we use a probability distribution in momentum space, where different wavelengths contribute to the generation of correlations across all spatial scales present in the system. 

The natural candidate for this distribution is the power spectrum, as proposed originally in Ref. \cite{GS1}. Since the power spectrum is the Fourier transform of the two-point correlation function, it naturally contains information about emergent spatial structures.
The radial configurational entropy density of the field fluctuations, $\Delta\phi\equiv\phi - \bar\phi$, is defined in the continuum as
\ee{
\frac{\dv S_C[\Delta\phi]}{\dv k} = -\frac{1}{\Omega_{d}}\intg{d}{\Omega} f(\vc k)\log f(\vc k),
}
where $\Omega_d$ is the solid angle in Fourier space and the modal fraction, $f(\vc k)$, is constructed from the normalized power spectrum of the field,
\ee{
f(\vc k) = \frac{\mathcal P(\vc k)}{\int_{\vc k'}\mathcal P(\vc k')}.
}
This is equivalent to Refs. \cite{GS1,GD1,GD2,GN} since
\ee{
\mathcal P(\vc k) =\mathcal{F}\langle\Delta\phi'\Delta\phi\rangle = |\mathcal F \phi (\vc k) |^2,
}
\noindent
where $\mathcal{F}$ denotes the Fourier transform operator: the modal fraction is the normalized Fourier transform of the two-point correlation function of the field fluctuations.
As defined, the modal fraction is a normalized probability distribution for the field fluctuations, with high-power modes being more likely than low-power modes. As such, it plays a similar role as the probability of a symbol in an alphabet in Shannon's theory, where now the symbol is a momentum mode. The total configurational entropy is then calculated by integrating (summing, for a discrete lattice, see Ref. \cite{GD2}) the CE density over all modes. Being computed from the normalized Fourier transform of the two-point correlation function, the CE is sensitive to spatial correlations at different distances, thus providing an informational measure of spatial coherence in the system. It is also a measure of the compressibility of the field, as it quantifies the average information in momentum space necessary to describe the field's state: a small number of more probable modes will decrease the CE or, equivalently, increase the field's compressibility.

\image{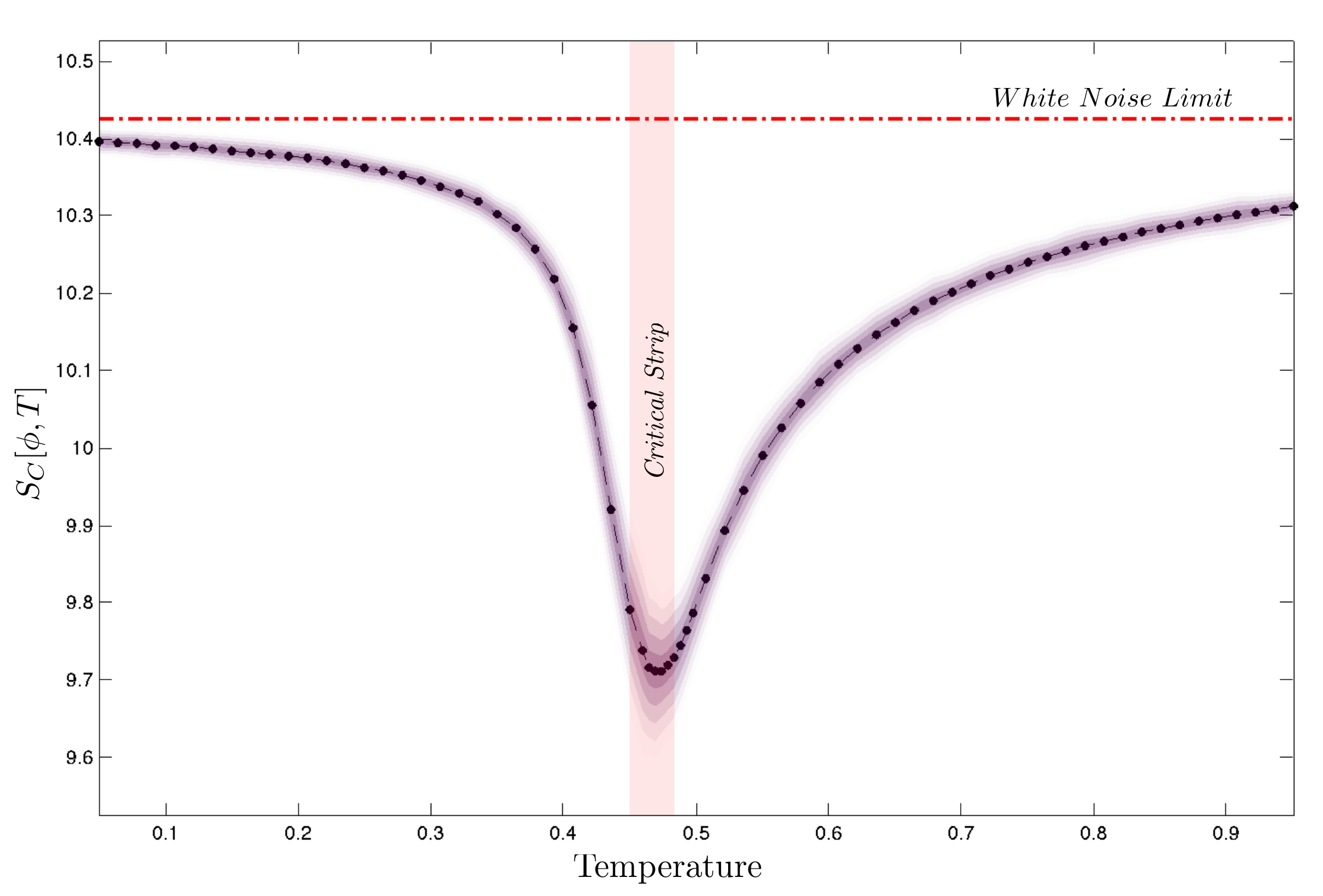}{The ensemble CE plotted as a function of temperature with 5$\sigma$ error bands. The horizontal line denotes the white noise limit for our lattice.}{CEvT}

\section{Results}

For each temperature $T$ we calculated the CE and its density. Results for the CE are plotted in 
Fig. \ref{CEvT}. The CE drops sharply in the critical strip where the phase transition occurs. 
The well-known increased spatial coherence at larger scales as criticality is approached translates into a larger departure from the maximum entropic state (``white noise limit'' line) characterized by a scale-free regime, as we show below. The critical point maximizes information about the spatial organization of the system. Compressibility is at a maximum, given that a few long wavelengths dominate the probability distribution: increased spatial coherence is equivalent to increased compressibility. In other words, minimizing CE implies in minimizing shape complexity.

The maximal (stored) information at criticality is equivalent to a reduction of uncertainty in the available phase space for the evolution of the system.  The CE can thus be interpreted as a measure of \textit{information storage} within the context of phase transitions. 

\begin{figure}
	\includegraphics[width=7 in]{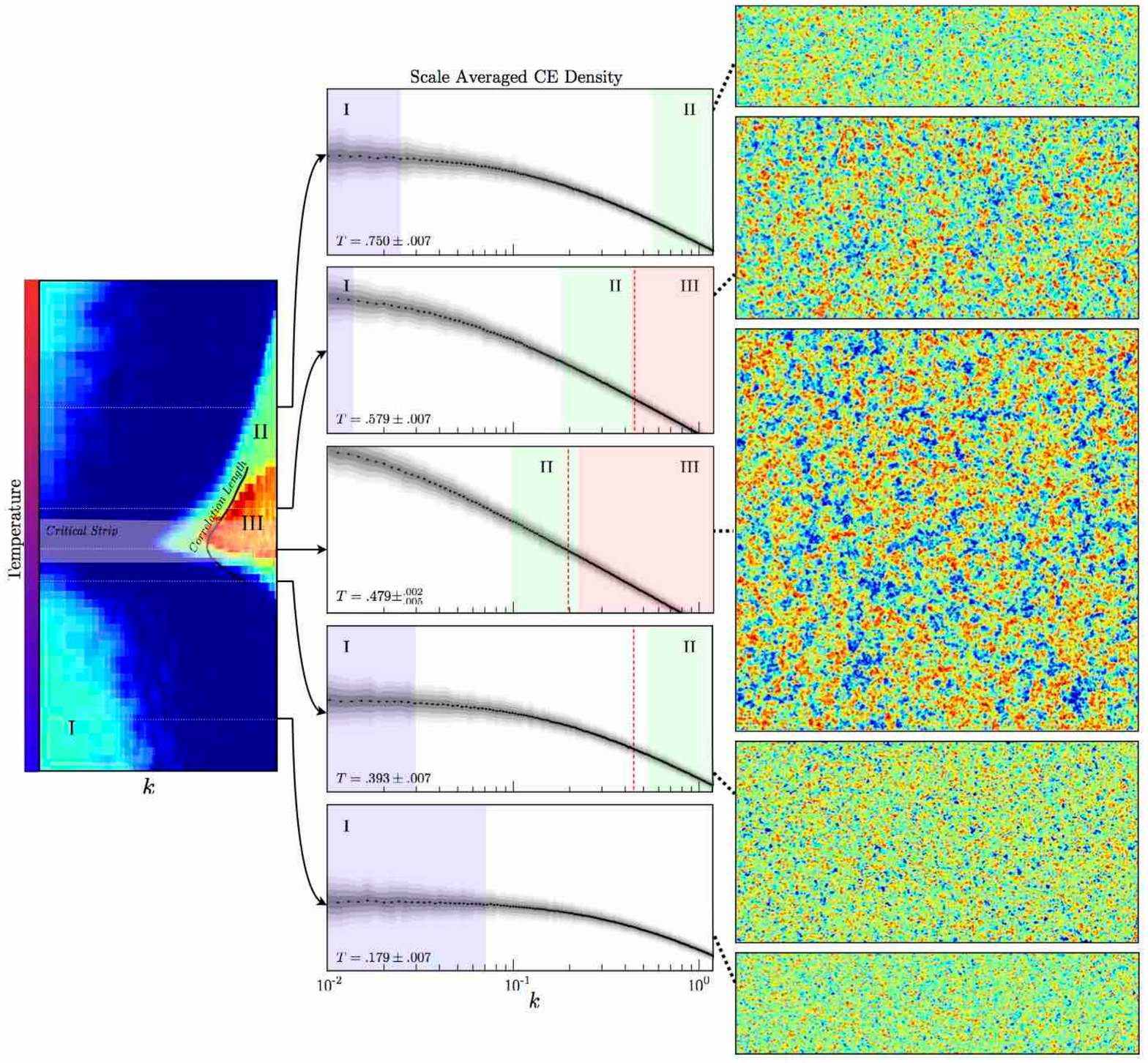}
	\caption{For different temperatures, there are three regions with different scaling behaviors: scale free (I), turbulent (II), and critical (III). At large scales the scale-free spectrum dominates. As the temperature approaches the critical temperature, the critical region moves from small to large scales. Between these two regions is a turbulent region. In the top panel one can examine the distribution of configurational entropy density at many scales.  Close to criticality, the correlation length diverges, leading to large correlated domains. Small scales become critical.}
	\label{fieldImages}
\end{figure}

To explore information transfer between modes as the system cycles through different equilibrium states, we investigate the behavior of the CE density at different temperatures.  As can be seen from Fig. \ref{fieldImages} there are 3 regimes where different scaling laws, of the form $k^{-\sigma}$, hold:

\begin{enumerate}[label={\Roman*}.]
\item \textbf{Scale Free} ($\sigma = 0$): away from criticality and for low-$k$ modes;
\item \textbf{Turbulent} ($\sigma = 5/3$): at length scales close to the correlation length (denoted by a dashed vertical line);
\item \textbf{Critical} ($\sigma = 7/4$): at or very near criticality and for scales smaller than the correlation length. 
\end{enumerate}

These regions shed light on the scales at which information about the various momentum modes and their relative roles is being stored and transferred, as we discuss next.

\section{Discussion}

\subsection{Scale-Free Regime}
The scale-free region only applies for low wave numbers, which correspond to large spatial scales. As we have seen, in information theory the uniform probability distribution maximizes entropy. It carries no information, in that it predicts all outcomes are equally likely: the default position when nothing is known about a system. Similarly, the scale-free region implies that the uniform modal fraction at these scales is storing no information about the fluctuations. At different temperatures, fluctuations at the largest spatial length-scales, where the scale-free regime holds, are uncorrelated.

To get a quantitative grasp on this regime, we have to look no further than white noise. Consider a $2d$ Gaussian random field, $\langle\xi\rangle=0,\langle\xi\xi'\rangle=\sigma^2\delta(x-x')$. The power spectrum is uniform, and the modal fraction is just the reciprocal of the momentum space volume. Since we are dealing with a discretized system, this volume is the number of cells in momentum space after cutting radial wave numbers larger than one-eighth of the Nyquist frequency, $\frac{\pi}{4}(N/4)^2$, leading to a CE of $\approx 10.85$. Numerics associated with the FFT introduce an additive correction to this analytical result, which a finite-scaling analysis reveals to be lattice-size independent. The CE of numerical white noise in our lattice is therefore $\approx 10.442\pm .004$, denoted by the horizontal dashed line in Fig. \ref{CEvT}. 

At low temperatures, the CE approaches the white noise limit, indicating that fluctuations get more and more uncorrelated at all scales. This seems reasonable since at low temperatures the correlation length is small and fluctuations are uncorrelated at most spatial scales. There is a departure from pure white noise, though, given that at any finite temperature there is a probability that some correlations at low wave numbers will be excited, a trend that grows as criticality is approached from below and from above. 

Going back to the alphabet analogy, a system far from criticality is nearly incompressible. Equipower modes take the same number of bits to encode, so there isn't a more efficient way of representing them. The informational narrative far from criticality is then one of ignorance: the state of the system is nearly that of white noise. 

\subsection{Turbulent Regime}

Searching for scaling laws as the temperature was increased, we found a surprising regime which obeys a power law identical to Kolmogorov turbulence in fluids \cite{Kolmogorov}. Unlike normal turbulence, which describes how an energy input at large spatial scales gets transported to smaller and smaller scales until it is dissipated away at the viscous length scale, the {\it informational turbulence} encountered here is a bottom-up phenomenon. Due to the nonlinear interactions of the order parameter and the free-energy cost of field gradients, small correlated domains tend to join together to become larger and larger. However, this trend cannot run away to organize the entire system because the scale-free fluctuations flowing into the system from the heat bath act to break them apart.  

Dimensional analysis, generalized from Kolmogorov's original argument, can be used to find this scaling law analytically. Within our standard formalism, CE is an information-theoretic measure with no units. We restore physical units by multiplying the CE by the Boltzmann constant, and dividing it by the mass of the system: $s_C=\frac{k_B}{M}S_C$. Note the units of this CE entropy-density: $\left[\frac{\dv s_C}{\dv k}\right]=\left[\text{Length}\right]^3 \left[\text{Time}\right]^{-2} \left[\text{Temperature}\right]^{-1}$. Following Kolmogorov we assume that the flow of energy between scales per unit mass, $\dot\epsilon$, is scale-independent. This energy is supplied by the heat bath and flows to larger scales. Since in the thermodynamic limit the only other dimensionful constant is the critical temperature, we posit an {\it ansatz} of the form:
\ee{
\frac{\dv s_C}{\dv k} = A T_C^\alpha \dot\epsilon^\beta k^\gamma,
}
where $A$ is a dimensionless proportionality constant. Consistency demands that $\alpha = -1,\ \beta=2/3, \ \gamma = -5/3$, and we obtain the scaling law in the turbulent regime. 

As criticality is approached, a resonant cooperation between deterministic and stochastic forces in the system drives modes to larger and larger spatial scales. As we have argued above, this is equivalent to an increase in information storage.
The turbulent regime determines the boundary of this resonant behavior, thus
defining an information-flow boundary between the scale-free (minimal information) and the critical regime (maximal information) spliced by correlation-length sized fluctuations (vertical dashed line in Fig. \ref{fieldImages}). 

\subsection{Critical Regime}

At criticality the effective temperature-dependent potential becomes quartic, with no curvature at the minimum. The corresponding divergence in the correlation length implies that fluctuations separated by large distances can exchange information, leading to larger correlated domains. Interpreting the CE as a measure of information storage, information is thus being maximally stored in these low-$k$ ordered structures. In the infinite-volume limit, this scaling law would cover all modes ($k\rightarrow 0$). 

To compute the critical scaling law recall the expressions for the two-point correlation function and power spectrum close to criticality,
\ee{
\langle\phi\phi'\rangle\propto \frac{e^{-r/r_c}}{(r/r_c)^{\eta}}&\xrightarrow{r_c\rightarrow\infty} \left(\frac{r}{r_c}\right)^{-\eta};
}
\ee{
\mathcal P(k) \propto \frac{r_c^{d-1}\sin\left[(d-1-\eta)\arctan kr_c\right]}{k(1+k^2r_c^2)^{\frac{d-1-\eta}{2}}}&\xrightarrow{r_c\rightarrow\infty} \frac{(kr_c)^{\eta}}{k^d},
}
where $\eta$ is the critical exponent, while $r_c$ is the correlation length, which diverges at criticality. The CE density at $T_C$ can then be expanded, with $\eta=\frac{1}{4}$ and $d=2$ \cite{Goldenfeld}, as
\ee{
\frac{\dv S_C}{\dv k} \propto k^{-7/4}+\mathcal O(k^{-7/4}\log k).
}
The ratio between the first term to the second goes as $\log r_c$ close to criticality. We can thus neglect the second term and obtain the scaling law in the critical regime \cite{GD2}. This scaling persists away from (but near) $T_C$, though suppressed to smaller and smaller spatial scales. 

Since this regime obtains only close to criticality and CE is minimized in the critical strip, we conclude that the modes obeying this scaling law are storing the majority of the information encoded in the system. Our numerical calculations show that modes obeying the critical scaling are comparable to or smaller than the correlation length, while turbulent modes are larger. The presence of these modes causes a drop in the CE of the system, which we interpret as an increase in its compressibility. 

\section{Summary}

In this paper we have examined an informational measure known as configurational entropy (CE) at a phase transition in a Landau model. We showed that the presence of three scaling regimes, their effect on the CE density, and the behavior of CE at criticality shed light on how phase transitions can be interpreted as storing and processing information. 

A scalar field coupled to a heat bath can be thought of in terms of computation. The field and its temporal derivatives at an initial time step are the input, or message. This message is processed via a channel to the next time step, the output being the time-evolved field. The channel consists of two parts: a deterministic evolution generated by the equations of motion, and a stochastic evolution (noise) generated by the heat bath. Information, in the traditional sense of reduction in uncertainty, is the property of the system to constrain output from input. Since we are concerned with the intrinsic information in a system at equilibrium, we must look at the fluctuations about the mean field to understand the system's computational properties.

We argued that the scalar field's CE is a measure of its compressibility as a message. The more compressible, the less random. The less random, the more its future is constrained by its present. Systems far from criticality are less compressible. Hence, as we approach criticality the system has a more determined future. This translates into the longevity of coherent structures at criticality. 

Although our investigation of the CE of the Landau model at equilibrium has elucidated the informational properties of critical phenomena, it has also opened up many questions. Apart from extending the current treatment to nucleation in first-order transitions, it would be interesting to investigate the role of topological defects in the informational narrative. Since these tend to have large gradients, their presence overwhelms the correlations in fluctuations. Disentangling the two in a way that preserves their informational properties would be a natural extension of the current work. 

The informational narrative of phase transitions presented here sheds light on the computational properties of systems at criticality. Our work has shown that the storage and processing of information is maximized at criticality as large and long-lived structures form, limited only by scales where the utilization of that information becomes turbulent. We hope that our approach points toward a clearer connection between statistical physics and complexity theory, in particular under the general prism of information theory. 

\acknowledgments
DS is supported by a Hull Fellowship at Dartmouth College. MG is supported in part by a Department of Energy grant DE-SC001038.\\

\end{document}